\documentclass{CVIS}

\usepackage{amsmath,bm}
\usepackage{graphicx}
\usepackage{enumerate}
\usepackage{comment}
\usepackage[numbers]{natbib}

\begin{document}

\title{SAMSON: Spectral Absorption-fluorescence Microscopy System for ON-site-imaging of algae}

\author{
\begin{tabularx}{\textwidth}{p{3.5cm} p{5.5cm} p{8cm}}
Jason L. Deglint \textsuperscript{[1]} & \textsuperscript{[1]} Vision and Image Processing Lab & \textsuperscript{[2]} School of Environmental Science and Engineering \\
Lyndon Tang \textsuperscript{[1]}  & Systems Design Engineering & Yat-sen University\\
Yitian Wang \textsuperscript{[1]}  & University of Waterloo & Guangzhou 510275, PR China\\
Chao Jin \textsuperscript{[1,2]}  & Ontario, Canada & \\
Alexander Wong \textsuperscript{[1]}  &  & \\
\end{tabularx}
}

\maketitle

%%%%%%%%%%%%%%%%%%%%%%%%%%%%%%%%%%%%%%%%%%%
\begin{abstract}
\vspace{-0.15in}
This paper presents SAMSON, a \textbf{S}pectral \textbf{A}bsorption-fluorescence \textbf{M}icroscopy \textbf{S}ystem for \textbf{ON}-site-imaging of algae within a water sample.  Designed to be portable and low-cost for on-site use, the optical sub-system of SAMSON consists of a mixture of low-cost optics and electronics, designed specifically to capture both fluorescent and absorption responses from a water sample.  The graphical user interface (GUI) sub-system of SAMSON was designed to enable flexible visualisation of algae in the water sample in real-time, with the ability to perform fine-grained exposure control and illumination wavelength selection.  We demonstrate SAMSON's capabilities by equipping the system with two fluorescent illumination sources and seven absorption illumination sources to enable the capture of multispectral data from six different algae species (three from the Cyanophyta phylum (blue-green algae) and three from the Chlorophyta phylum (green algae)).  The key benefit of SAMSON is the ability to perform rapid acquisition of fluorescence and absorption data at different wavelengths and magnification levels, thus opening the door for machine learning methods to automatically identify and enumerate different algae in water samples using this rich wealth of data.
\vspace{-0.15in}
\end{abstract}

%%%%%%%%%%%%%%%%%%%%%%%%%%%%%%%%%%%%%%%%%%%
\section{Introduction}
\vspace{-0.15in}
Harmful algae blooms (HABs) occur when different types of cyanobacteria and algae grow out of control in a water body due to a combination of different factors.
These HABs not only have a catastrophic effect on marine wildlife and recreational water use, but more importantly on water treatment plants and drinking water quality as they have shown to produce lethal toxins that can cause diarrhea, vomiting, blistered mouths, dry coughs, and headaches~\cite{falconer1996potential}.
It is of utmost importance that these algae blooms are monitored carefully in order to better understand their behaviour by measuring a number of different factors, one of which is the algae type and quantity.

The standard practice to determine which types of algae are in a water body and their associated quantities is a multi-step manual process that can be described as follows.  First, a water sample is transported to a remote facility, which requires that it be preserved during transportation and thus is often a costly and time-consuming process.  Secondly, once the sample arrives and has been prepared, a trained taxonomist will manually identify the different species present by observing it through an eyepiece attached to a microscope, which is a very straining, tedious, and time-consuming task.  Aside from the time-consuming nature of this manual process, Culverhouse \textit{et al.} have shown that the average taxonomist hit rate is between 67\% - 83\%, with a high variability between different taxonomists~\cite{culverhouse2003experts}.  Furthermore, Clerck \textit{et al.} demonstrated that the number of algae species is rapidly increasing while at the same time the number of algae taxonomists is decreasing~\cite{clerck2013algal}.
As a result of these challenges associated with the traditional approach for algae type identification and quantification, novel methods and techniques for doing so in a more rapid, consistent, and low-cost manner is highly desired.

\begin{figure}[!t]
    \centering
    \includegraphics[width=0.72\columnwidth]{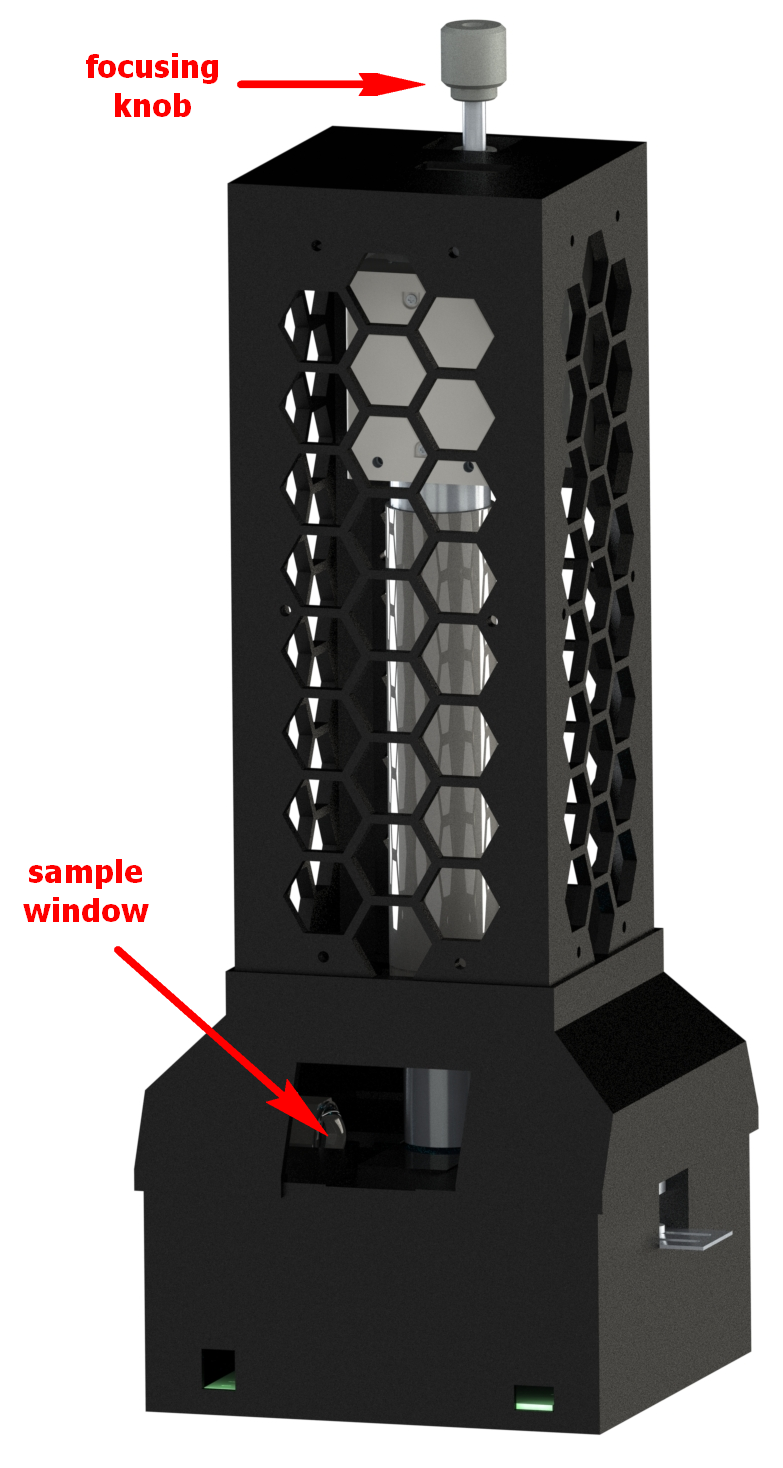}
    \caption{The proposed SAMSON system. The user places the water sample through the sample window and adjust the focus with the focusing knob.  All control over the illumination sources and sensor of SAMSON is done through the graphical user interface (Fig.~\ref{fig:ui}).}
    \label{fig:3d_print}
    \vspace{-0.2in}
\end{figure}

One approach to addressing these challenges is to design instruments that are capable of collecting reliable data in a rapid manner and then leverage pattern recognition and machine learning methods to perform automated identification and quantification, as recommended by Walker \textit{et al.}~\cite{walker2000image}. 
Motivated by this, in this paper we attempt to tackle this instrument design portion of this approach by introducing a low-cost, portable imaging system capable of collecting high-quality data in a rapid manner directly on-site.  More specifically, the proposed \textbf{S}pectral \textbf{A}bsorption-fluorescence \textbf{M}icroscopy \textbf{S}ystem for \textbf{ON}site-imaging (SAMSON), as shown in Figure~\ref{fig:3d_print}, is specifically designed to rapidly collect both absorption and fluorescent imaging data of different algae within a water sample.
As a result, SAMSON enables the rapid collection of large amounts of insightful data directly on-site that can then be leveraged by different machine learning methods to automatically identify algae types and associated quantities. Additionally, the collected data can be viewed directly from a large computer screen instead of a traditional eye-piece, meaning that a taxonomist can observe data more effectively, store data for later use, and share data among colleagues.

%%%%%%%%%%%%%%%%%%% FIGURES %%%%%%%%%%%%%%%%%%% 

\begin{figure}[!t]
    \centering
    \includegraphics[width=.88\columnwidth]{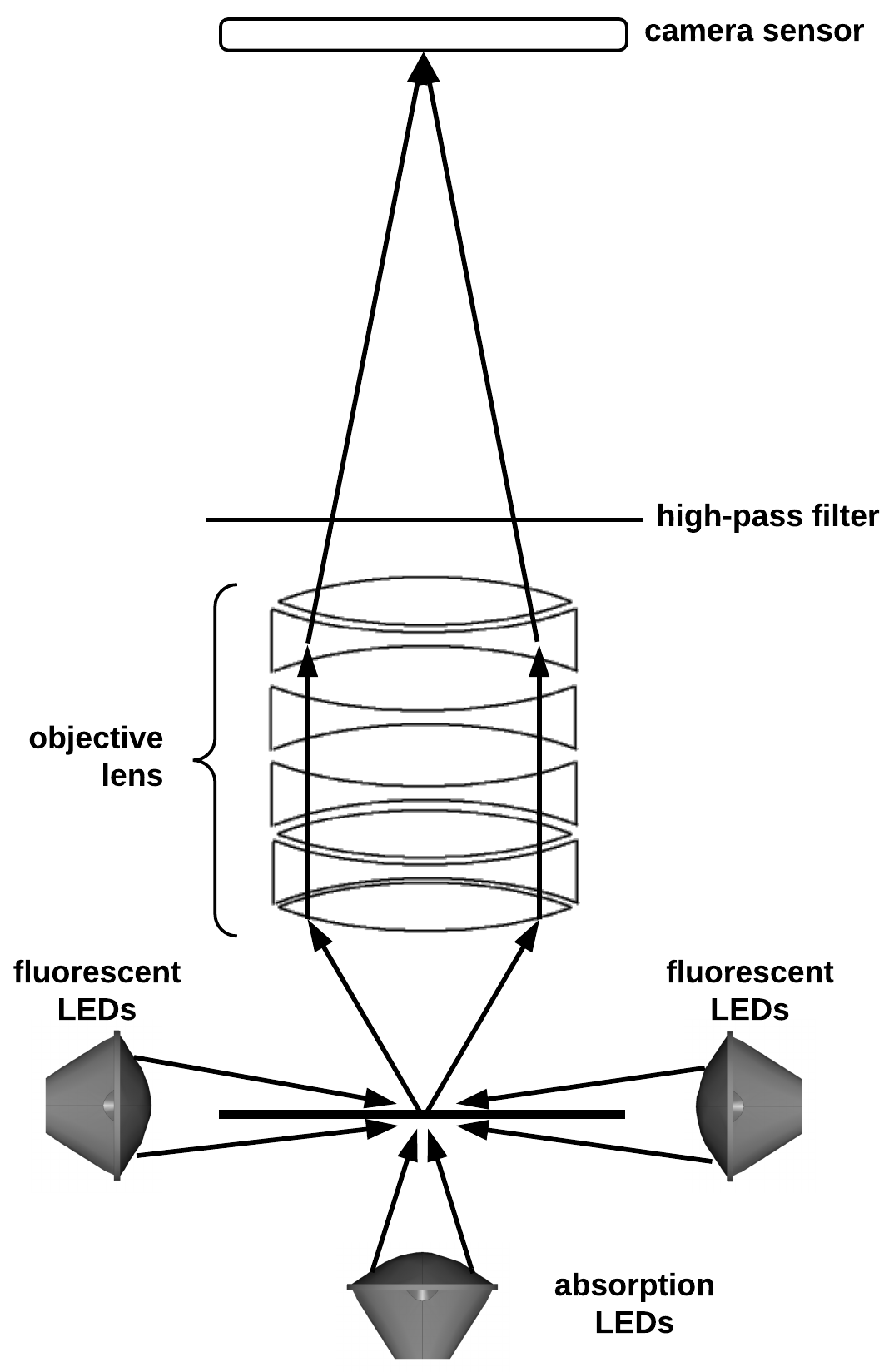}
    \caption{The optical sub-system of SAMSON captures multi-wavelength absorption-fluorescence images of water samples.}
    \label{fig:functional_system}
        \vspace{-0.2in}
\end{figure}

%%%%%%%%%%%%%%%%%%%%%%%%%%%%%%%%%%%%%%%%%%%
\section{Methodology}
\vspace{-0.15in}
To tackle the current challenges associated with water monitoring, which is not only time-consuming and costly, but also requires long delays due to the need to transport water remotely for manual analysis, the proposed SAMSON system is designed to be low-cost and highly portable, enabling such as system to be stationed directly at the water source for rapid, on-site inspection of a given water sample.  SAMSON consists of two main sub-systems: i) an optical sub-system of SAMSON for rapid  collection of both absorption and fluorescent imaging data (as detailed in Section~\ref{sec:hw}), and ii) a graphical user interface (GUI) for rapid visualisation of collected imaging data and adjustment of illumination and sensor parameters (as detailed in Section~\ref{sec:sw}).  SAMSON is designed specifically to be modular in nature; that is, it has the ability to house different combinations of illumination sources and different lenses.  This was done to allow for the ability to conduct rapid experiments with different illumination wavelength combinations, for both the absorption and fluorescent illumination sources, as well as at different magnification levels so that the optimal configuration may be found.

%%%%%%%%%%%%%%%%%%%%%%
\subsection{Optical sub-system} \label{sec:hw}
    \vspace{-0.1in}
The SAMSON optical sub-system is shown in Figure~\ref{fig:functional_system} and can be described as follows.  First, light is emitted from either the fluorescent illumination sources or from the absorption illumination sources onto the slide that contains a water sample.  If the fluorescent illumination source is active, it will cause the algae to auto-fluorescence. If the absorption illumination source is active, the algae will either absorb or transmit the light.  The emitted or transmitted light then passes through a series of lenses (acting together as an objective magnification lens) as well as a high-pass-filter, and finally onto an imaging sensor.  A vertical translation stage which controls the fine tune focusing of the image, is connected to the sensor and a focusing knob extends from the top of SAMSON allowing a user to focus the image. 

%%%%%%%%%%%%%%%%%%%%%%
\subsection{Graphical User Interface sub-system} \label{sec:sw}
\begin{figure}[!b]
    \centering
    \includegraphics[width=\columnwidth]{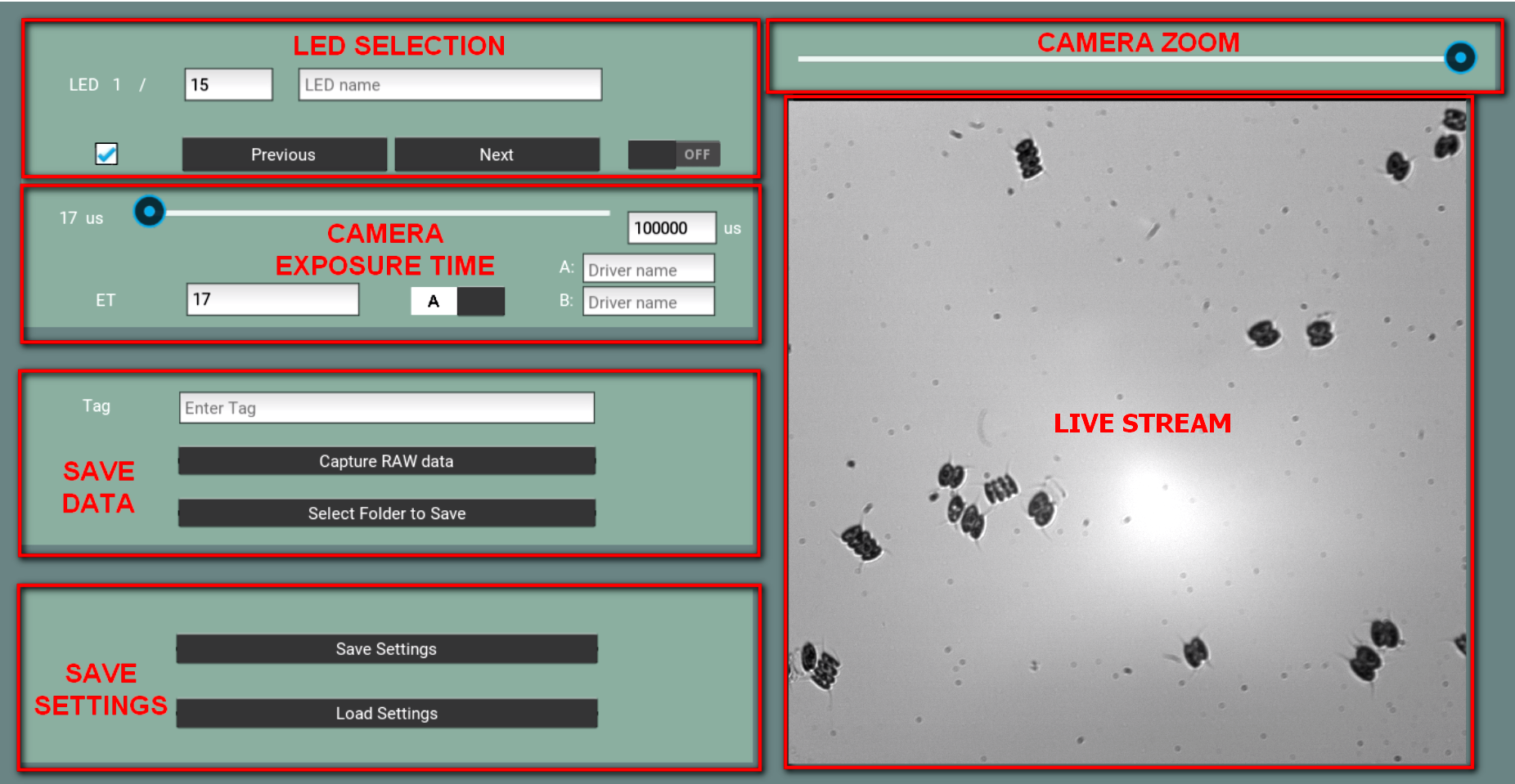}
    \caption{The SAMSON GUI sub-system enables the flexible selection of different illumination sources and changes in exposure time of the sensor, all while viewing the water sample in real-time.}
    \label{fig:ui}
\end{figure}

\vspace{-0.1in}
The SAMSON graphical user interface (GUI) sub-system is shown in Figure~\ref{fig:ui}.  In a standard scientific environment, multiple software programs are necessary for the purpose of image acquisition to control the various components (e.g., sensor, individual illumination sources, etc.).  As such, this typically required the user to switch between programs during the image acquisition process, which is not only time-consuming but also prone to error.  Therefore, the SAMSON GUI sub-system is designed to provide all required functions in one concise and user-friendly interface, thus drastically improving the efficiency of the image acquisition process. The three main functions in the GUI sub-system are: i) live-stream visualisation, ii) sensor exposure time control, and ii) illumination source selection and control.  All operations can be performed in the GUI sub-system through sliders.

Furthermore, to increase the reliability of data, it is necessary to perform calibration during the image acquisition acquisition process. Based on the observation of microorganisms, we have found that they can move rapidly in water sample, thus causing inconsistencies and motion artefacts to exhibit in the acquired imaging data.  To mitigate this issue, the back-end of the GUI sub-system leverages software triggers to greatly accelerate the image acquisition speed of SAMSON by precisely timing and synchronising the illumination sources at different wavelengths with the sensor acquisition process, thus increasing the quality and consistency of the data captured especially under rapid motion by the microorganisms being imaged.

\begin{figure*}[]
    \centering
    \includegraphics[width=.95\textwidth]{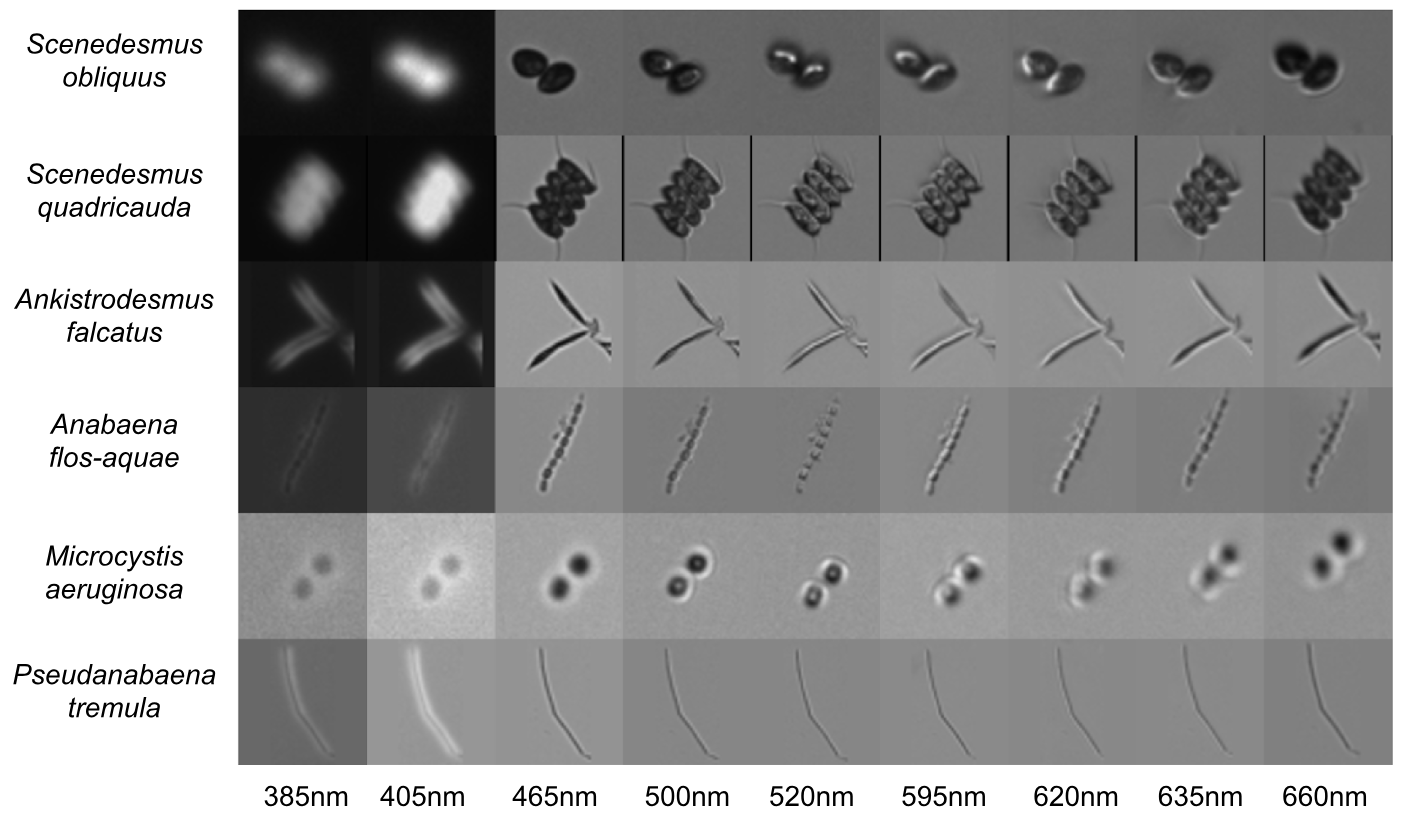}
    \caption{Six algae samples from two phyla groups were imaged using SAMSON, which was configured to capture absorption and fluorescent at nine different wavelengths. From the Cyanophyta phyla (blue-green algae): \textit{Microcystis aeruginosa} (CPCC 300), \textit{Anabaena flos-aquae} (CPCC 067), and \textit{Pseudanabaena tremula} (CPCC 471). From the Chlorophyta phyla (green algae): \textit{Scenedesmus obliquus} (CPCC 005), \textit{Scenedesmus quadricauda} (CPCC 158), and \textit{Ankistrodesmus falcatus} (CPCC 366).}
    \label{fig:cropped_images}
\end{figure*}

%%%%%%%%%%%%%%%%%%%%%%%%%%%%%%%%%%%%%%%%%%%
\section{Experimental Setup \& Results}
\vspace{-0.15in}
To demonstrate the capabilities and efficacy of SAMSON for the purpose of on-site imaging of algae, the system was configuring to capture absorption and fluorescent at nine different wavelengths, and subsequently used to image six different types of algae in this study. In Section~\ref{sec:setup_algae}, we discuss which six algae were chosen in this study. Since SAMSON is modular in nature, in Section~\ref{sec:setup_hw} we discuss which wavelengths and magnification optical components were selected for SAMSON in this study.

%%%%%%%%%%%%%%%%%%%%%%
\subsection{Algae Samples} \label{sec:setup_algae}
\vspace{-0.1in}
For this study, six different types of algae from the Canadian
Phycological Culture Centre (CPCC) were chosen to demonstrate the imaging capabilities of the proposed SAMSON system.  
These six algae types, as well with their respective CPCC number, are broken into there respective phyla and are described as follows:

\begin{enumerate}[I.]
\setlength{\itemsep}{0pt}
\setlength{\parskip}{0pt}
    \item Cyanophyta (blue-green algae or cyanobacteria)
    \begin{enumerate}[1.]
        \item \textit{Microcystis aeruginosa} (CPCC 300)
        \item \textit{Anabaena flos-aquae} (CPCC 067)
        \item \textit{Pseudanabaena tremula} (CPCC 471)
    \end{enumerate}
    \item Chlorophyta (green algae)
    \begin{enumerate}[1.]
    \setcounter{enumii}{3}
        \item \textit{Scenedesmus obliquus} (CPCC 005)
        \item \textit{Scenedesmus quadricauda} (CPCC 158)
        \item \textit{Ankistrodesmus falcatus} (CPCC 366)
    \end{enumerate}
\end{enumerate}

Three species from the Cyanophyta phyla were chosen since cyanobacteria are known to produce lethal toxins during a harmful algae bloom. Both \textit{Microcystis sp.} and \textit{Anabaena sp.} have both been shown to produce toxins during bloom conditions~\cite{falconer1996potential}.

%%%%%%%%%%%%%%%%%%%%%%
\subsection{Optical sub-system configuration} \label{sec:setup_hw}
\vspace{-0.1in}
For this study, the optical sub-system of SAMSON is configured with LEDs at nine different wavelengths acting as absorption and fluorescent illumination sources to demonstrate the capabilities of rapid acquisition of absorption and fluorescent signals of algae samples.
Of the LEDs chosen, two were used as fluorescent illumination sources and seven where used as absorption illumination sources. The exact wavelengths, along with their associated colours, are described as follows.
The fluorescent wavelengths used were 385 nm (ultraviolet) and 405 nm (ultraviolet).
The absorption wavelengths used were 465 nm (blue), 500 nm (cyan), 520 nm (green), 595 nm (amber), 620 nm (red-orange), 635 nm (red), and 660 nm (deep-red).
The ultraviolet LEDs were chosen as fluorescent illumination sources since all algae are known to contain Chlorophyll-a, as they use photosynthesis to convert energy from the sun~\cite{barsanti2014algae}.
Although 385 nm and 405 nm are not at the peak excitation wavelength of Cyanophyta or Chlorophyta, it is sufficient to induce auto-fluorescence in both phyla.

Further, in this study, a series of lenses act together as a 20x objective magnification lens, and a Point Grey Grasshopper 3 CMOS camera, with a resolution of 4.1 mega-pixels and a pixel pitch of 5.5 um, was used as the sensor. This resulted in a spatial resolution of 0.26 um/pixel as measured by the USAF 1951 calibration chart. When accounting for the relative brightness of each LED, the effective frame-rate of SAMSON as configured for this study was approximately 30 frames-per-second when acquiring all nine absorption and fluorescent acquisitions.

%%%%%%%%%%%%%%%%%%%%%%%%%%%%%%%%%%%%%%%%%%%
\subsection{Results \& Discussion}
\vspace{-0.1in}
The resulting absorption and fluorescent imaging cube across all nine wavelengths for one sample of each species can be seen in Figure~\ref{fig:cropped_images}.
One significant observation is that all three species of the Cyanophyta phyla have significantly more fluorescence at 385 nm and 405 nm, as seen in Figure~\ref{fig:cropped_images}.
This matches Poryvkina \textit{et al.} findings~\cite{poryvkina2000analysis}, and indicates that such spectral information could be used when building a machine learning algorithm to automatically differentiate and count different types of algae.
This initial observation indicates that using pattern recognition and machine learning methods could likely differentiate between different phyla groups.

Our hypothesis is that such a device as SAMSON, which provides both spectral (fluorescent and absorption) and morphological data, will allow for both low variability and high accuracy when using machine learning algorithms to build an on-site real-time monitoring system for algae, especially since different algae groups are known to have different pigments~\cite{barsanti2014algae}. As discussed by Li \textit{et al.}, multiple artificial intelligence approaches, such as machine vision, pattern recognition and machine learning algorithms are a cost-effective and labour-reducing solution to microorganism classification~\cite{li2017survey}.

%%%%%%%%%%%%%%%%%%%%%%%%%%%%%%%%%%%%%%%%%%%
\section{Conclusions}
\vspace{-0.15in}
In this work we presented a \textbf{S}pectral \textbf{A}bsorption-fluorescence \textbf{M}icroscopy \textbf{S}ystem for \textbf{ON}site-imaging (SAMSON), designed to enable on-site data collection on water samples for algae.
The main advantage of SAMSON is the ability of rapid acquisition of high quality fluorescence and absorption data.
Having this stream of high-quality data opens the door for using machine learning methods for automatic identification and enumeration of algae samples.
\vspace{-0.15in}
%%%%%%%%%%%%%%%%%%%%%%%%%%%%%%%%%%%%%%%%%%%
\section*{Contributions}
\vspace{-0.15in}
JLD, CJ, and AW conceived and designed the SAMSON system.
JLD and LT designed the electrical sub-system and 3D model while LT implemented the design.
JLD and YW designed the graphical user interface (GUI) while YW implemented the design.
JLD collected and processed the data.
All authors contributed to the writing of this manuscript.

%%%%%%%%%%%%%%%%%%%%%%%%%%%%%%%%%%%%%%%%%%%
\section*{Acknowledgements}
\vspace{-0.15in}
The authors would like to thank Heather Roshen at the Canadian Phycological Culture Centre (CPCC) for preparing the algae samples and Velocity Science for providing tools and resources for proper data collection.
This research was funded by the Natural Sciences and Engineering Research Council of Canada (NSERC) and Canada Research Chairs program.

%%%%%%%%%%%%%%%%%%%%%%%%%%%%%%%%%%%%%%%%%%%
\bibliographystyle{unsrt}
\vspace{-0.15in}
\setlength{\bibsep}{0.1pt}
\bibliography{main.bib}
\vspace{-0.15in}
\end{document}